# CONTENT BASED VIDEO RETRIEVAL

B. V. Patel[1] and B. B. Meshram[2]

[1] Shah & Anchor Kutchhi Polytechnic, Mumbai, INDIA
patelbv@acm.org
[2]Computer Technology Department, VJTI, Mumbai, INDIA
bbmeshram@vjti.org.in

## ABSTRACT

*Content based video retrieval is an approach for facilitating the searching and browsing of large image collections over World Wide Web. In this approach, video analysis is conducted on low level visual properties extracted from video frame. We believed that in order to create an effective video retrieval system, visual perception must be taken into account. We conjectured that a technique which employs multiple features for indexing and retrieval would be more effective in the discrimination and search tasks of videos. In order to validate this claim, content based indexing and retrieval systems were implemented using color histogram, various texture features and other approaches. Videos were stored in Oracle 9i Database and a user study measured correctness of response.*

## 1. INTRODUCTION

During recent years, methods have been developed for retrieval of videos based on their visual features. Color, texture, shape, motion and spatial-temporal composition are the most common visual features used in visual similarity match. Realizing that inexpensive storage, ubiquitous broadband Internet access, low cost digital cameras, and nimble video editing tools would result in a flood of unorganized video content; researchers have been developing video search technologies for a number of years. The recent trends in video creation and delivery technology have brought the need for such tools to the forefront. Video retrieval continues to be one of the most exciting and fastest growing research areas in the field of multimedia technology [1].

Despite the sustained efforts in the last years, the paramount challenge remains bridging the semantic gap. By this we mean that low level features are easily measured and computed, but the starting point of the retrieval process is typically the high level query from a human[3]. Translating or converting the question posed by a human to the low level features seen by the computer illustrates the problem in bridging the semantic gap. However, the semantic gap is not merely translating high level features to low level features. The essence of a semantic query is to understand the meaning behind the query. This can involve understanding both the intellectual and emotional sides of the human, not merely the distilled logical portion of the query but also the personal preferences and emotional sub tones of the query and the preferential form of the results. Content-based image retrieval (CBIR), also known as query by image content (QBIC) and content-based visual information retrieval (CBVIR) is the application of computer vision to the video retrieval problem, that is, the problem of searching for video in large databases.

 "Content" in this context might refer to colors, shapes, textures, or any other information that can be derived from the image itself. Without the ability to examine video content, searches must rely on metadata such as captions or keywords, which may be laborious or expensive to produce. The demand for intelligent processing and analysis of multimedia information has been rapidly growing in recent years[12]. Among them, content-based video retrieval is the most challenging and important problem of practical value. It can help users to retrieve desired video from a large video database efficiently based on the video contents through user interactions. The video retrieval system can be roughly divided into two major components: a module for extracting representative features from video frames and one defining an appropriate similarity model to find similar video frames from database. Many approaches used different kinds of features to represent a video frame, including color histogram, shape information, texture, text analysis. Few approaches integrate features to improve the retrieval performance [8].

Our proposed approach consists of various modules for key frame extraction, indexing, features extraction, similarity search etc. We use a dynamic programming approach to compute the similarity between the feature vectors for the query and feature vectors in the feature database. Proposed Video Storage and Retrieval System, stores and manages a large number of video data and allows users to retrieve videos from the database efficiently. It is interactive web based application which takes video frame from users and retrieve the information from the database. Database consists of various video data like still video frames, audio and video. The retrieval is based on the content of the video object.

Proposed System provides different functionality for two main clients-which are Administrator and user. Administrator is responsible for controlling the entire database including security and adding, updating and deleting videos to and from database. User can only retrieve videos based on submitted query based on content as well on metadata. Rest of the paper is organized as follow: Section two deals with the System Analysis and Design, in section three we discuss system design, section four describes proposed algorithms, we present experimental results in section five and finally section six concludes.

## 2. SYSTEM ANALYSIS AND DESIGN

In this section we present system analysis and design of proposed systems. System analysis model defines user requirements, establishes basis of system design and defines set of validation requirements needed for testing implemented system. System design is the technical kernel of System engineering and is applied regardless of the system process model that is used. Beginning once system requirements have been analyzed and specified, System design is the first of three technical activities—design, code generation, and test—that are required to build and verify the system. Each activity transforms information in a manner that ultimately results in validated proposed System.

### 2.1. System Analysis

At the core of the model lies the data dictionary—a repository that contains descriptions of all data objects of system. Three different diagrams surround the core. The entity relation diagram (ERD) depicts relationships between data objects. The ERD is the notation that is used to conduct the data modeling activity. The attributes of each data object noted in the ERD can be described using a data object description.

The data flow diagram (DFD) serves two purposes- to provide an indication of how data are transformed as they move through the system and  to depict the functions and sub functions that transform the data flow.

The DFD provides additional information that is used during the analysis of the information domain and serves as a basis for the modeling of function. A description of each function presented in the DFD is contained in a process specification (PSPEC). The state transition diagram (STD) indicates how the system behaves as a consequence of external events. To accomplish this, the STD represents the various modes of behavior (called states) of the system and the manner in which transitions are made from state to state. The STD serves as the basis for behavioral modeling.

## 2.2. *ER Diagram*

The title is to be written in 20 pt. Garamond font, centred and using the bold and "Small Caps" formats. There should be 24 pt. (paragraph) spacing after the last line.

The object/relationship pair is the cornerstone of the data model. These pairs can be represented graphically using the entity/relationship diagram. A set of primary components are identified for the ERD data objects, attributes, relationships, and various type indicators. The primary purpose of the ERD is to represent data objects and their relationships. ER diagram shows the relationship between all the entities in the system. An Entity can be described using a number of attributes. The Relationship shows how all the entities interact with each other. There are two basic entities in proposed system, which are as shown in figure 1.

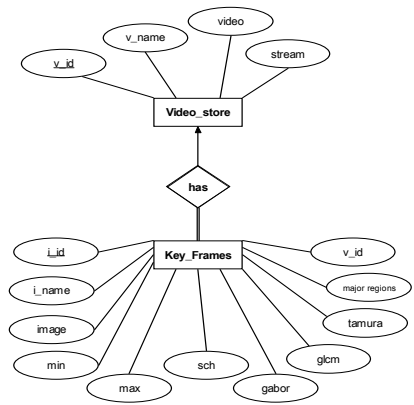

Figure 1. ER Diagram

Attributes:   Key_frames(i_id,i_name,image,min,max,sch,gabor,glcm,tamura,majorRegions,v_id)
Video_store (v_id,v_name,video,stream)

## 2.3. Use Case Diagram

A use-case is a scenario that describes how System is to be used in a given situation. To create a use-case, the analyst must first identify the different types of people (or devices) that use the system or product. These actors actually represent roles that people (or devices) play as the system operates. An actor is anything that communicates with the system or product and that is external to the system itself. A use case corresponds to a sequence of transaction, in which each transaction is invoked from outside the system (actors) and engages internal objects to interact with one another and with the system surroundings. Use case represents specific flows of events in the system; it is also a graph of actors, a set of use cases enclosed by a systems boundaries communication between actors and the use cases. It is shown in following figure 2.

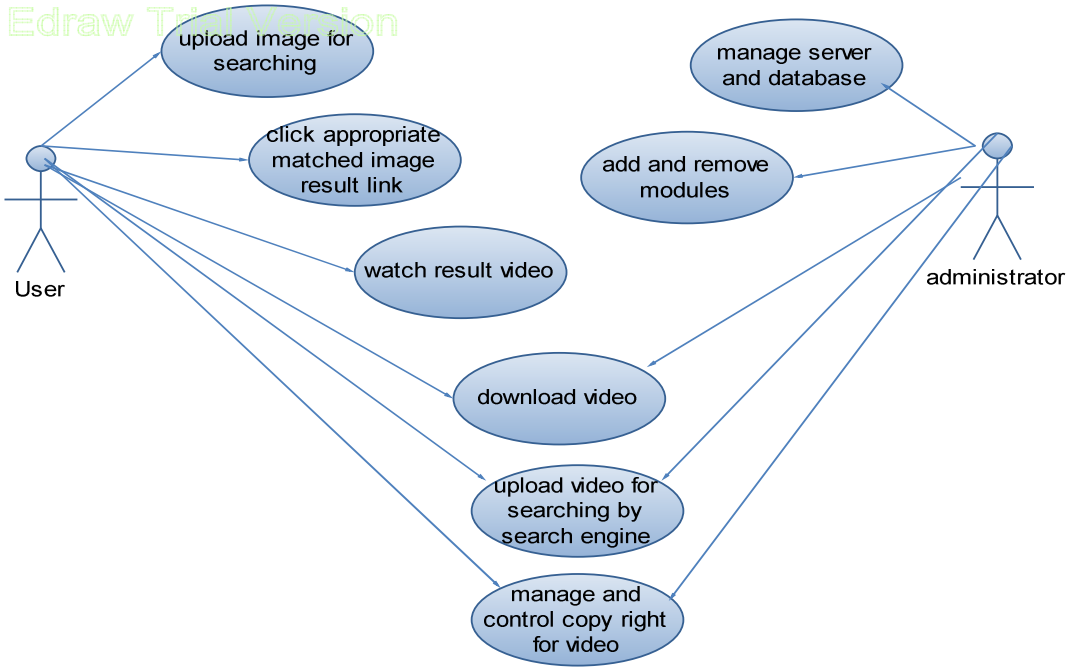

Figure 2. Use case diagram

## 2.4. Data Flow Diagram

As information moves through System, it is modified by a series of transformations. A data flow diagram is a graphical representation that depicts information flow and the transforms that are applied as data move from input to output. The data flow diagram may be used to represent a system or System at any level of abstraction. In fact, DFDs may be partitioned into levels that represent increasing information flow and functional detail. Therefore, the DFD provides a mechanism for functional modeling as well as information flow modeling. Figure 3 demonstrates DFD for proposed systems.

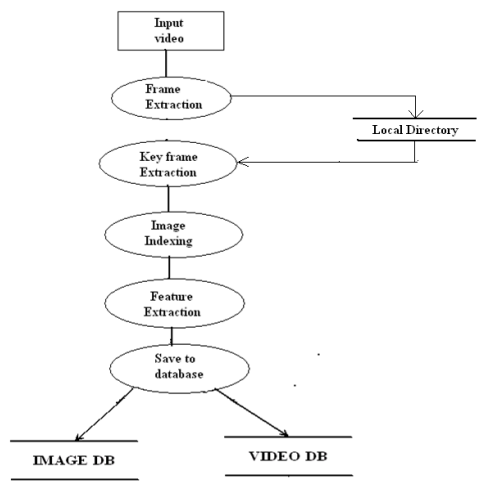

Figure 3. DFD for proposed systems

## 3. SYSTEM DESIGN

Each of the elements of the analysis model provides information that is necessary to create the design models required for a complete specification of design. System requirements, manifested by the data, functional, and behavioral models feed the design task. Using one of a number of design methods, the design task produces a data design, an architectural design, an interface design, and a component design.

### 3.1. Block Diagram of Proposed System

The System contains two main modules. The Administrator is responsible for addition, deletion and modification of the multimedia objects. User can use only the search engine to search image, audio or video as shown in figure 4.

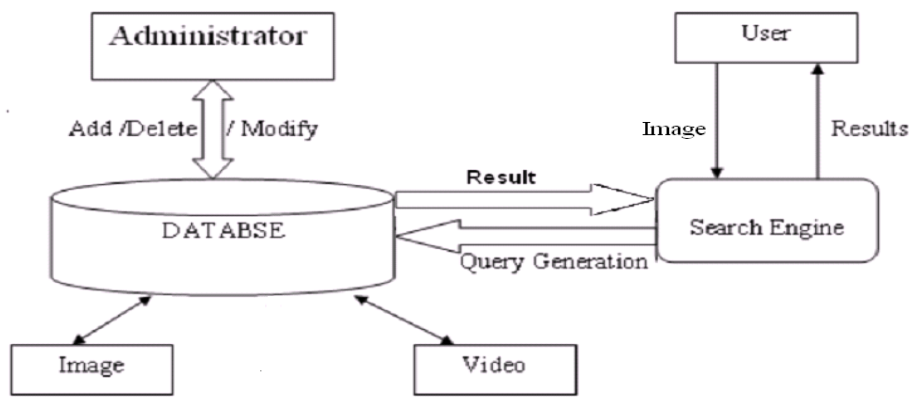

Figure 4. System Block Diagram

### 3.2. Component Diagram

It models the physical components in design the way of looking at the components is the packages . Apackage is used to show how we can group together classes , which in essence are smaller scale components .

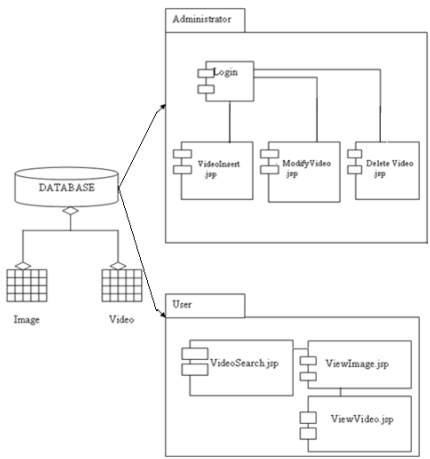

Figure 5. Component diagram

Here administrator manages login, addition, modification and deletion of video from database and user searches video based on index frame and views similar videos as shown in figure 5.

## 3.3. Deployment Diagram

Deployment Diagram shows the configuration of runtime processing elements and the software components, processes and objects that live in them. It is shown in figure 6. A software component instance represents runtime manifestations of code units. In most cases, component diagrams are used in the conjunction with deployment diagram to show how physical modules of code are distributed on various hardware platforms. In many cases, component and deployment diagrams can be combined. A deployment diagram is graph of nodes connected by communication association. Nodes may contain a component instance, which means that component lives or runs at that mode. Components may contain objects; this indicates that the object is a part of component. Components are connected to other components by dashed arrow dependencies, usually through interfaces, which indicate one component uses service of another. Each node or processing element in the system represented by a three dimensional box. Connection between the node (or platforms) themselves are shown by solid lines.

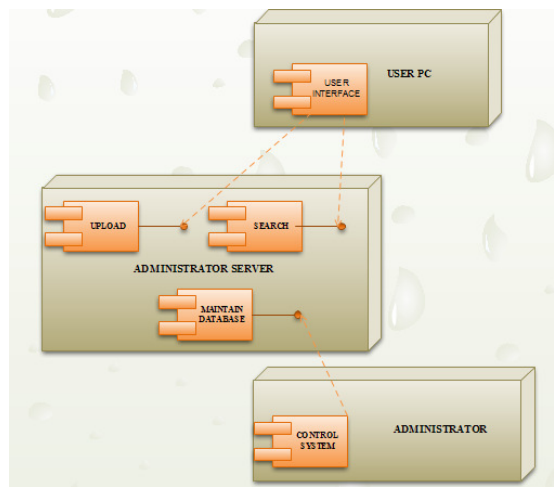

Figure 6. Deployment Diagram

## 3.4. Database Design

Database design is used to define and then specifies the structure of business objects used in the client/server system. The analysis required to identify business objects is accomplished using business process engineering methods. Conventional analysis modeling notation such as the ERD, can be used to define business objects, but a database repository should be established in order to capture the additional information that cannot be fully documented using a graphic notation such as an ERD.

In this repository, a business object is defined as information that is visible to the purchasers and users of the system, not its developers. This information, implemented using a relational database, can be maintained in a design repository. Following are the database tables used for implementing proposed systems on which queries can be performed.

Video_store(v_id, v_name, video,stream)

Key_frames(v_id,i_name,image,min,max,sch,gabor,glcm,tamura,majorRegions,v_id)

Where v_id - video id,

       v_name - name of video file,

       video - video file,

       stream - stream of keyframes,

       i_id – video frame id,

       min-max - indexing range,

       sch – simple colour histogram,

       gabor, glcm, tamura - feature texture string,

       majorRegion - no. of max regions

Using following Oracle query   Video_store table is created for indexing and retrieval of video.

```
CREATE TABLE  "VIDEO_STORE"
   (     "V_ID" NUMBER NOT NULL ENABLE,
         "V_NAME" VARCHAR2(60),
         "VIDEO" ORD_ Video,
         "STREAM" BLOB,
         "DOSTORE" DATE,
         PRIMARY KEY ("V_ID") ENABLE
   )
```

Using following Oracle query  KEY_FRAMES table is created for storing key frames and their features.

```
CREATE TABLE  "KEY_FRAMES"
   (     "I_ID" NUMBER NOT NULL ENABLE,
         "I_NAME" VARCHAR2(40) NOT NULL ENABLE,
         "IMAGE" ORD_ Image,
         "MIN" NUMBER,
         "MAX" NUMBER,
         "SCH" VARCHAR2(1500),
         "GLCM" VARCHAR2(250),
         "GABOR" VARCHAR2(1500),
         "TAMURA" VARCHAR2(500),
         "MAJORREGIONS" NUMBER,
         "V_ID" NUMBER,
         PRIMARY KEY ("I_ID") ENABLE
   )
```

# 4. ALGORITHMS PROPOSED AND USED

## 4.1 Key frame extraction Algorithm

It works on group of frames extracted from a video. It takes a list of files or frames in order in which they will be extracted. It is based on predefined threshold that specify whether 2 video frame are similar. It s main function is to choose smaller number of video representative key frames.

Starts from $1^{st}$ frame from sorted list of files.

If consecutive frames are within threshold, then two frames are similar.

Repeat process till frames are similar, delete all similar frames & take $1^{st}$ as key -frame.

Start with next frame which is outside threshold & repeat the steps i to a for all frames of video .

**Input -** Frames of video extracted by video to jpeg converter.

**Output -** Key frames with less similarity & representing video.

**Pseudo code for key frame extraction**

1] Let i & j be integer variables, ri1 & ri2 be RenderedImage objects.
2] Get all Jpeg files in files array.
3] Initialise length to length of files array.
4] Do
a] i=0
      b] while i is not equal to length
           i] Create RenderedImage ri1 object that hold rescaled IVersion of image file i.
           ii] Do
                  a] j=i+1
                  b] while j is not equal to length
                  c] Create RenderedImage ri2 object that holds rescaled IVersion of
image file j.
                  d] dist = difference between ri1 & ri2.
                  e] if(dist > 800.0)
                        { i=j-1; break; }
                else
                     delete file j.
                  f] j=j+1
      c] i=i+1
5] End

## 4.2 Histogram Based Range Finder Indexing Algorithm

We implemented tree based indexing algorithm consisting of grouping frames into pixel ranges such as 0-255 on first level, 0-127 or 128-255 on second level so on third, fourth level. Algorithm level by level calculates min – max range finally if when thresholding criteria is not satisfied, frame is grouped onto previous level into min –max range. Every frame is part of first level as it satisfies thresholding criteria.

1. Calculate histogram of image.
2. Calculate min–max range by grouping pixel count.
3. Find if thresholding criteria are satisfied.
4. Go to next lower level & repeat step from 2 to 4.
5. If criteria are not satisfied, image is grouped on previous level where min – max range is satisfied as shown in figure 7.

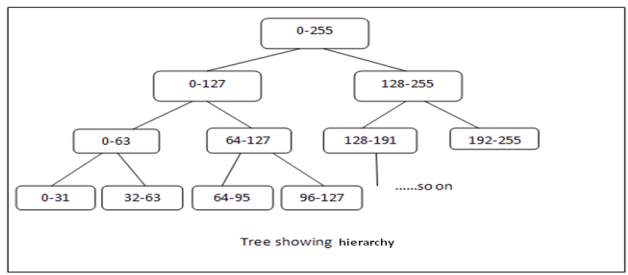

Figure 7. Indexing tree

**Indexing:**

**Input :** Video frame after choosing key frames as image.

**Output :** min – max range that groups image in particular group.

**Pseudo code for indexing process :**

Variables min & max store range for indexing process.

1] Let min = 0,max = 255;

2] Let long sum = 0;

3] Let double result = 0;

4] //1st block test

      A] i=min

      B] while i is not equal to max

            i] sum = sum + histogram[i]

      C] result = sum/900.0;

      D] if(result>55.0)

              {min = 0;max = 127;}

                else

              {min = 128;max = 255;}

      E] i=i+1

5] //2nd block test

      sum = 0, result = 0;

6] if(min = = 0 && max = = 127)

      {

    A] i=0

    B] while i is not equal to 63

          i] sum = sum + histogram[i]

    C] result = sum/900.0;

    D] if(result>60.0)

           {min = 0;max = 63;}

      else

      {

                I] sum = 0, result = 0;

          II] for (int i  =  64; i < 127; i++)

                a] sum = sum + histogram[i];

                III] result = sum/900.0;

                IV] if(result>60.0)

                {min = 64;max = 127;}

      }

    }

7] sum = 0, result = 0;

8] if(min = = 128 && max = = 255)

      {

    A] for (int i  =  128; i < 191; i++)

          i] sum = sum + histogram[i];

    B] result = sum/900.0;

    C] if(result>60.0)

          {min = 128;max = 191;}

      else

      {

                I] sum = 0, result = 0;

          II] for (int i  =  192; i < 255; i++)

                a] sum = sum + histogram[i];

                III] result = sum/900.0;

                IV] if(result>60.0)

                {min = 192;max = 255;}

      }

    }

```
 9] //3rd block test
        sum = 0, result = 0;
10] if(min = = 0 && max = = 63)
          {
        A] for (int i  =  0; i < 31; i++)
                          i) sum = sum + histogram[i];
        B] result = sum/900.0;
        C] if(result>60.0)
                          {min = 0;max = 31;}
             else
             {
                          I] sum = 0, result = 0;
                     II] for (int i  =  32; i < 63; i++)
                          a] sum = sum + histogram[i];
                     III] result = sum/900.0;
                          IV] if(result>60.0)
                          {min = 32;max = 63;}
                }
             }

11] sum = 0, result = 0;
12] if(min = = 64 && max = = 127)
          {
        A] for (int i  =  64; i < 95; i++)
                          i]sum = sum + histogram[i];
        B] result = sum/900.0;
        C] if(result>60.0)
                          {min = 64;max = 95;}
             else
                          {
                  I] sum = 0, result = 0;
                  II] for (int i  =  96; i < 127; i++)
                          a]sum = sum + histogram[i];
                          III] result = sum/900.0;
                          IV] if(result>60.0)
                          {min = 96;max = 127;}
                           }
             }

13] sum = 0, result = 0;
14] if(min = = 128 && max = = 191)
          {
        A] for (int i  =  128; i < 159; i++)
                          i]sum = sum + histogram[i];
        B] result = sum/900.0;
        C] if(result>60.0)
                          {min = 128;max = 159;}
                  else
                          {
                  I] sum = 0, result = 0;
                  II] for (int i  =  160; i < 191; i++)
                          a] sum = sum + histogram[i];
                  III] result = sum/900.0;
                  IV] if(result>60.0)
                          {min = 160;max = 191;}
```

```
                                }
            }

15] sum = 0, result = 0;
16] if(min = = 192 && max = = 255)
            {
        A] for (int i  =  192; i < 223; i++)
                        a]sum = sum + histogram[i];
        B] result = sum/900.0;
        C] if(result>60.0)
                        {min = 192;max = 223;}
                else
                        {
                        I] sum = 0, result = 0;
        II] for (int i  =  224; i < 255; i++)
                        a] sum = sum + histogram[i];
                        III] result = sum/900.0;
                        IV] if(result>60.0)
                        {min = 224;max = 255;}
            }
        }
```

## 4.3.  GLCM Texture Feature Extraction

The Gray Level Co-occurrence Matrix1 (GLCM) and associated texture feature calculations are image analysis techniques. Given an image composed of pixels each with an intensity (a specific gray level), the GLCM is a tabulation of how often different combinations of gray levels co-occur in an image or image section.

**Input :** image as PlanarImage object
**Output** : string containing feature values.
**Pseudo code for GLCM texture:**
public GLCM_Texture(PlanarImage im, boolean preprocess):
1]  If(preprocessor is true) input = preprocess the image in im;
2] otherwise input = im;
3] create a data structure Raster for input image
4] initialize k=0 ,h=o ,w=0;
        i] while h is not equal to height of image
                a] while i is not equal to width of image ;
                A] Initialise pixels[w][h] with value for pixel in image.
                B] Initialise pixel[k++] with value for pixel in image.

**private** PlanarImage preprocess(PlanarImage input)
{
**1] if** (input has IndexColorModel)
        A] Retrieve the IndexColorModel of image;
        B] Cache the number of elements in each band of the colormap in mapSize integer variable.
        C] Allocate an array for the lookup table data i.e. lutData = **new byte**[3][mapSize];
        D] Load the lookup table data from the IndexColorModel.
                i] Get Red in lutData[0] from IndexColorModel object ;
                ii] Get Green in lutData[1] from IndexColorModel object ;
                iii] Get Blue in lutData[2] from IndexColorModel object ;

    E] Create the lookup table object i.e. LookupTableJAI  using lutData array;
    F] Replace the original image i.e. input with the 3-band RGB image using lut object.

2] **if** ( NumBands in input > 1)
    A] Let matrix = {{ 0.114, 0.587, 0.299, 0 }};
    B] Replace the original image i.e. input with bandcombineb using matrix;
  **3]return** input;
public void calcTexture()
1] Let a = 0 be integer variables;
2] Let b = 0, y=0, x=0 be integer variables;
3] Let offset and i be integer variables;
4] Let glcm be a 2-D array of 257 by 257;
5] while y is not equal to height
    i]offset = y*width;
6] while x is not equal to width
    i] i = offset + x;
    ii] a = 0xff & pixel[i];
    iii] b = 0xff & pixels[x+*step*][y];
    iv] glcm [a][b] +=1;
    v] glcm [b][a] +=1;
    vi] pixelCounter +=2;
7] while a is not equal to 257
    i] while b is not equal to 257
        a] glcm[a][b]= glcm[a][b] / pixelCounter;
8] while a is not equal to 257
    i] while b is not equal to 257
        a] asm= asm+ (glcm[a][b]*glcm[a][b]);
9] Initialize double px= 0;
10] Initialize double py= 0;
11] Initialize double meanx= 0.0;
12] Initialize double meany= 0.0;
13] Initialize double stdevx= 0.0;
14] Initialize double stdevy= 0.0;
15] while a is not equal to 257
    i] while b is not equal to 257
        a] px= px+a*glcm [a][b];
        b] py= py+b*glcm [a][b];
16] while a is not equal to 257
    i] while b is not equal to 257
        a] stdevx= stdevx+(a-px)*(a-px)*glcm [a][b];
        b] stdevy= stdevy+(b-py)*(b-py)*glcm [a][b];
17] while a is not equal to 257
    i] while b is not equal to 257
        a] correlation= correlation+ (a-px)*(b-py)*glcm [a][b]/(stdevx*stdevy);
18] while a is not equal to 257
    i] while b is not equal to 257
        a] IDM=IDM+ glcm[a][b]/(1+(a-b)*(a-b))
19] while a is not equal to 257
    i] while b is not equal to 257
        a] if (glcm[a][b]==0) { }
            b] else
            entropy = entropy - glcm[a][b] * *log*(glcm[a][b]) ;

public String getStringRepresentation()
1] Build new object StringBuilder sb;

2] Append the string GLCM texture to sb;
3] Append pixelCounter to sb;
4] Append asm to sb;
5] Append contrast to sb;
6] Append correlation to sb;
7] Append IDM to sb;
8] Append entropy to sb;
9] Return string format of sb;

## 4.4. Gabor Texture Feature

Gabor Feature is a linear filter used for edge detection. Frequency and orientation representations of Gabor filters are similar to those of the human visual system, and they have been found to be particularly appropriate for texture representation and discrimination.

**Input :** image as PlanarImage object
**Output** : string containing feature values.
**pseudo code for Gabor Texture Feature Extraction:**
public double[] getFeature(BufferedImage image)
1] Get buffered image and store image to max height;
2] Create Raster object which calls the getRaster() using variable image;
3] Define an array GrayLevel which stores width & height of buffered image;
4] Define a temporary array;
5] For (int i = 0; i <Width of imageRaster; i++)
    i] For (int j = 0; j < Height of imageRaster; j++)
        a] Store pixel values into GrayLevel array;
6] Define a 1-dimensional vector FeatureVector;
7] Define a 2-dimentional vector magnitude which stores the magnitude of grayLevel array;
8] Declare the variable ImageSize to store the value of  Width * Height;
9] Define a 2-dimentioal vector MagnitudeForVariance;
10] If (gaborWavelet == null)
    i] Call precomputeGaborWavelet using grayLevel array;
11] For (int m = 0; m < $M$; m++)
    i] For (int n = 0; n < $N$; n++)
        a] Calculate featureVector[m * $N$ + n * 2] = magnitudes[m][n] / imageSize;
        b] For (int i = 0; i < length of magnitudesForVariance; i++)
            A] For (int j = 0; j < length of magnitudesForVariance[0]; j++)
                I] MagnitudesForVariance[i][j] = 0;
        c] For (int x = $S$; x < Width of image; x++)
        I] For (int y = $T$; y < Height of image; y++)
        a] Compute magnitudesForVariance[m][n] +=
Math.$pow$(Math.$sqrt$(Math.$pow$(this.gaborWavelet[x - $S$][y - $T$][m][n][0], 2)
+ Math.$pow$(this.gaborWavelet[x - $S$][y - $T$][m][n][1], 2))
 - featureVector[m * $N$ + n * 2], 2);
            II] Calculate featureVector[m * $N$ + n * 2 + 1]
=Math.$sqrt$(magnitudesForVariance [m][n]) / imageSize;
12] Set this.gaborWavelet to null;
13] Return featureVector;

## 4.5 Color histogram Extraction

The color space of frame is quantizes into a finite number of discrete levels. Each of this level becomes bin in the histogram. The color histogram is then computed by counting the number of theses discrete levels. Color Histogram(h) is then computed with the color information like $h_r(i)$, $h_g(i)$, $h_b(i)$ to represent the color domain.

> **Input :** image as BufferedImage object
> **Output** : string containing histogram values.
> **pseudo code for color histogram :**
> public void extract(BufferedImage image)
> 1] Let pixel be 2-D array of size width & height.
> 2] Create a data structure Raster for input image.
> 3] x = 0 , y=0 & while x is not equal to width of image
>     A] while y is not equal to height of image
>         i) pixel[x][y]= corresponding value in raster.
>         ii) **if**(histogramType != HistogramType.*HMMD*) histogram[quant(pixel)]++.
>     public String getStringRepresentation()
> 1] Create a StringBuilder object.
> 2] Append histogramType to sb.
> 3] Append  ' ' to sb.
> 4] Append length of  histogram to sb.
> 5] Append  ' ' to sb.
> 6] for (int i = 0; i < length of  histogram. i++)
>     A] Append histogram[i] to sb.
>     B] Append  ' ' to sb.
> 7] Return string format of sb.

## 4.6. Superficial (native) similarity Algorithm

This approach is to extract image signature with 25 representative pixels, each in R, G, B. For each of 25 locations over image take 5 * 5 matrix & find mean pixel value for matrix. Using this process vector array with 25 mean values for image signature is created.

> **Input :** images as BufferedImage object
> **Output** : difference between image signatures.
> **pseudo code for Superficial (naive) similarity descriptor:**
> private RenderedImage rescale(RenderedImage i)
>
> 1] float scaleW = 300, scaleH =300;
>     // Scales the original image
> 2] Create ParameterBlock pb = new ParameterBlock();
> 3] Adding input image to pb
> 5] Adding height required to pb
> 6] Adding filter InterpolationNearest for scaling to pb
> 7] Return scaled image
>
> private Color averageAround(RenderedImage i, double px, double py)
>
> 1] Create  RandomIter  object named as iterator.
> 2] let pixels & accum be 1-D arrays.
> 3] Let sampleSize = 15, numPixels = 0.
> 4] for (double x = px * *baseSize* - sampleSize; x < px * *baseSize* + sampleSize; x++)
>     I] for (double y = py * *baseSize* - sampleSize; y < py * *baseSize* + sampleSize; y++)
>         A] Get pixel value at x,y location.
>         B] accum[0] += pixel[0].
>         C] accum[1] += pixel[1].
>         D] accum[2] += pixel[2].
>         E] numPixels++.

5) Average the accumulated values.
      accum[0] /= numPixels, accum[1] /= numPixels, accum[2] /= numPixels.
6) return Color object with accum[0], accum[1], accum[2] as RGB values.

public String getStringRepresentation()
1] StringBuilder sb = new StringBuilder(25 * 5);
2] append("NaiveVector") to sb
3] append(' ') to sb
4] for (int i = 0; i < length of signature; i++)
      I] for (int j = 0; j < length  of signature[i]; j++)
            A] append(signature[i][j]) to sb
            B] append(' ') to sb
5] Return string format of sb;

## 4.7 Autocorrelogram for color

A color correlogram expresses how the spatial correlation of pairs of colors changes with distance. A color histogram captures only the color distribution in an image and does not include any spatial correlation information.

  **Input :** images as BufferedImage object
  **Output** : string containing correlogram values.
  **pseudo code for autocorrelogram :**
public void extract (BufferedImage bi)
1] Get Raster r from bi object's raster producing method;
2] Create array histogram = new int[256];
3] for (int i = 0; i < length of histogram; i++)
      a] Initialise histogram[i] = 0;
4] Create array quantPixels = new int[width of raster][height of raster]
5] quantize colors for each pixel (done in HSV color space):
      Create array pixel = new int[3], hsv = new int[3];
6] for (int x = 0; x < width of raster ;x++)
      a] for (int y = 0; y < height of raster ;y++)
            I] converting to HSV:  convertRgbToHsv using pixel from raster r, hsv array
            II] quantize the actual pixel: quantPixels[x][y] = quantize hsv
            III]  for normalization: histogram[quantPixels[x][y]]++
7] Create array correlogram = new float[256][maxDistance];
8] for (int i1 = 0; i1 <length of correlogram; i1++)
      a] for (int j = 0; j < length of correlogram[i1]; j++)
            I] correlogram[i1][j] = 0;
9] Create array tmpCorrelogram = new int[maxDistance];
10] for (int x = 0; x < Width of rster r; x++)
      a] for (int y = 0; y < Height of rster r; y++)
            I] Create variable color = quantPixels[x][y];
            II] getNumPixelsInNeighbourhood(x, y, quantPixels, tmpCorrelogram)
            III] for (int i = 0; i < maxDistance; i++)
                A] correlogram[color][i] += tmpCorrelogram[i];
11]Create array max = new float[maxDistance] & initialise with 0
12]for (int c = 0; c < numBins; c++)
      a]for (int i = 0; i < maxDistance; i++)
            I] max[i] = Maximum betweencorrelogram[c][i] & max[i]
13] for (int c = 0; c < numBins; c++) {
      a] for (int i = 0; i < maxDistance; i++)
            I] correlogram[c][i] = correlogram[c][i] / max[i];

## 4.8. Simple Region Growing

Region growing is a simple region-based image segmentation method. It is also classified as a pixel-based image segmentation method. It is a classic stack-based region growing algorithm:

Find a pixel which is not labeled. Label it and store its coordinates on a stack.

While there are pixels on the stack, do:

Get a pixel from the stack (the pixel being considered).

Check its neighbors to see if they are unlabeled and close to the considered pixel; if are, label them and store them on the stack.

Repeat all steps until there are no more pixels on the image.

**Input :** images as BufferedImage object
**Output** : number of regions, number of holes , number of major regions.
**pseudo code for Simple Region Growing :**
**private** PlanarImage preprocess(PlanarImage input)
**1] if** (input has IndexColorModel)
    A] Retrieve the IndexColorModel in IndexColorModel icm object;
    B] Cache the number of elements in each band of the colormap in mapSize integer variable.
    C] Allocate an array for the lookup table data i.e.**byte**[][] lutData = **new**
            **byte**[3][mapSize];
    D] Load the lookup table data from the IndexColorModel.
          i] Get Red in lutData[0] from icm object ;
          ii] Get Green in lutData[1] from icm object ;
          iii] Get Blue in lutData[2] from icm object ;
    E] Create the lookup table object i.e. LookupTableJAI lut using lutData array;
    F] Replace the original image i.e. input with the 3-band RGB image using lut object.
2] **if** ( NumBands in input > 1)
    A] Let matrix = {{ 0.114, 0.587, 0.299, 0 }};
    B] create a new ParameterBlock;
    C] AddSource input image to pb;
    D] Add matrix to pb;
    E] Replace the original image i.e. input with bandcombineb;
3] Should we binarize it?
    A] create a new ParameterBlock;
    B] Add source input image to pb;
    C] Add sampling parameter to pb
    D] Add **new int**[]{256} to pb;
    E] Add **new double**[]{0} to pb;
    F] Add **new double**[]{256 to pb;
    G] Calculate the histogram of the image and its Fuzziness Threshold.
        Create PlanarImage dummyImage to hold histogram ;
    H] Histogram h = Property of histogram from dummyImage;
    J] double[] thresholds = h.getMinFuzzinessThreshold();
    K] create a new ParameterBlock;
    L] AddSource input image to pb;
    M] Add thresholds[0] to pb;
    N] Replace the original image i.e. input with bynarized;
    4] **Let** kernelMatrix =          { 0, 0, 0, 0, 0,

0, 1, 1, 1, 0,
0, 1, 1, 1, 0,
0, 1, 1, 1, 0,
0, 0, 0, 0, 0 };

5] Create the kernel object using the array kernelMatrix.

6] create a new ParameterBlock;

7] AddSource input image to p;

8] Add kernel to p;

9] Replace the original image i.e. input with dilated;

10] create a new ParameterBlock;

11] AddSource input image to p;

12] Add kernel to p;

13] Replace the original image i.e. input with erode;

14] create a new ParameterBlock;

15] AddSource input image to p;

16] Add kernel to p;

17] Replace the original image i.e. input with erode;

19] create a new ParameterBlock;

20] AddSource input image to p;

21] Add kernel to p;

22] Replace the original image i.e. input with dilated;

23] **return** input image object;

**public void** run()

1] numberOfRegions = 0;numhole=0;

2] Create a Stack of Point objects with mustDo as name;

3] for(int h=0;h<height;h++)

    A] for(int w=0;w<width;w++)

        I] position++;

        II] if (labels[w][h] < 0)

    `           a] if(pixels[w][h]==0)

                i] numhole++;

            b] numberOfRegions++;

            c] Add a new Point(w,h) to mustDo;

            d] labels[w][h] = numberOfRegions;

            e] put(numberOfRegions,1) into count;

        III] while(mustDo.size() > 0)

            a] Point thisPoint = mustDo.get(0); mustDo.remove(0);

            b]for(int th=-1;th<=1;th++)

                i] for(int tw=-1;tw<=1;tw++)

                    1)int rx = thisPoint.x+tw;

                    2)int ry = thisPoint.y+th;

                3)if ((rx < 0) || (ry < 0) || (ry>=height) || (rx>=width)) continue;

                    4)if (labels[rx][ry] < 0)

                A)if (pixels[rx][ry] == pixels[thisPoint.x][thisPoint.y])

                    i)mustDo.add(new Point(rx,ry));

                    ii)labels[rx][ry] = numberOfRegions;

                iii)count.put(numberOfRegions, count.get(numberOfRegions)+1);

4] position = width*height;

# 5. EXPERIMENTAL RESULTS

Java programming language is used for implementing proposed system. Ecllipse, Jcreator, Apache Tomcat server, Oracle 9i are used for the development of the system. Proposed systems is implemented for the java virtual machine enabled windows based pc's with internet connection. Sample videos and frames were downloaded from www.archive.org. We have considered different categories of images like e-learning, sports, cartoon, movies, etc. Table II presents the average precision values at the top 20, 30, 50, and 100 retrieved video based on various features. We chose these recall points comes from the fact that for retrieval, the system is able to present 25 or 30 thumbnails per page (or screen), and browsing a set of 60 or 100 key frames is not too painful for users (for instance, the well-known Google web retrieval system (http://www.google.com) presents by default 20 query results for images, and 10 query results for text).

Table 1. PRECISION AT 20, 30, 50 AND 100 DOCUMENTS.

|  | GLCM | Gabor | Tamura | Histogram | Autocorrelogram | Simple Region Grawing | Combined |
|---|---|---|---|---|---|---|---|
| Avg. prec.at 20 frames | 0.435 | 0.586 | 0.568 | 0.398 | 0.412 | 0.520 | 0.629 |
| Avg. prec.at 30 frames | 0.423 | 0.528 | 0.514 | 0.368 | 0.405 | 0.468 | 0.553 |
| Avg. prec.at 50 frames | 0.410 | 0.489 | 0.469 | 0.324 | 0.369 | 0.434 | 0.494 |
| Avg. prec.at 100 frames | 0.354 | 0.396 | 0.412 | 0.310 | 0.342 | 0.397 | 0.421 |

From Table I, we observe that our combined approach outperforms all the other methods. In short, our experimental results show that by combining various approaches to take advantage of different levels of representations, we achieve better retrieval performance over individual approach. Following figures 8 to figure 13 shows the input, intermediate results and output of proposed content based video retrieval systems. Due to limitation of space we are unable to list java functions and packages used for implementation.

## 5.1. Sample Input And Outputs

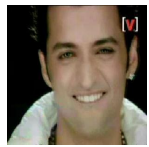

Figure 8. Input Query Image

Algorithm : SimpleColorHistogram
Output : min = 0, max=127
Histogram : RGB 256 19401 2570 1848 1098 774 552 425 312 231 214 169 176 186 152 174 157 149
128 128 125 126 136 118 131 130 110 141 125 134 133 150 138 148 139 134 142 154 163 135 177 168

180 188 213 231 223 231 215 215 221 227 233 214 231 220 222 239 223 236 236 239 264 255 226 267
344 350 381 457 443 446 434 526 512 546 544 504 530 575 633 532 552 545 578 547 511 502 521
499 465 520 572 588 596 513 597 582 537 490 548 516 520 523 552 562 610 567 592 624 631 601 699
695 804 828 929 819 841 729 631 623 490 462 454 431 423 377 393 335 369 393 347 334 409 413 543
521 623 588 550 356 335 274 202 184 166 158 129 146 136 126 123 117 117 110 87 92 101 107 115 112
133 154 158 137 160 170 154 135 141 154 179 159 157 150 155 127 132 163 149 168 194 204 233 255
212 225 210 208 195 163 186 124 153 126 123 138 120 139 92 110 96 95 95 64 86 97 81 96 109 104 98
88 89 79 72 46 46 36 40 36 31 26 26 30 15 16 16 14 13 12 13 6 18 9 15 16 7 11 10 10 8 6 6 7 5 4 0 3 2 0 1
5 0 0

Algorithm : GLCM_Texture
Output :
180000.0  0.03020352129629111  87.88990000000011  2.274446602930954E-4  0.5007650756513622
6.821227228133351

Algorithm : Gabor Texture
Output :
gabor  60  8.756868030900593  0.09356683213681918  8.759811592603452  0.0935982839918384
8.759811592603462      0.09359828399183825      9.04978747435506      0.09669666626207571
9.05282949891942  0.09672917018957598  9.05282949891942  0.096729170189576  9.041419889746448
0.09660725891005423      9.044459101623616      0.09663973278388098      9.044459101623616
0.09663973278388112      9.017710748889508      0.09635392755975444      9.02074199110002
0.0963863162779682      9.02074199110002      0.0963863162779682      8.950311139124567
0.09563376505979511      8.953319725414202      0.09566591170027569      8.953319725414202
0.09566591170027565      8.756868030900593      0.09356683213681918      8.759811592603452
0.0935982839918384  8.759811592603462  0.09359828399183825  0.0  0.0  0.0  0.0  0.0  0.0  0.0  0.0  0.0
0.0  0.0  0.0  0.0  0.0  0.0  0.0  0.0  0.0  0.0  0.0  0.0  0.0  0.0  0.0  0.0

Algorithm : Tamura Texture
Output :
Tamura 18 14620.0  44.24798484297803  1098.0  234.0  196.0  212.0  136.0  131.0  116.0  160.0  133.0  159.0
151.0  200.0  171.0  219.0  270.0  258.0

Algorithm : SimpleRegionGrowing
Output : Majorregions : 2

Algorithm : AutoColorCorrelogram
Output :
ACC  4  0.7046129  0.7100176  0.71587896  0.72141516  0.0011786222  0.0010277173  8.970384E-4
7.8307267E-4  0.07746393  0.074670404  0.07256122  0.07076817  0.008961593  0.008232119  0.0077413702
0.007344682  3.048161E-4  1.9723865E-4  1.5539247E-4  1.2421152E-4  0.0054460475 ,..............

Algorithm : NaiveVector
Output :
NaiveVector    java.awt.Color[r=0,g=0,b=0]    java.awt.Color[r=0,g=0,b=0]    java.awt.Color[r=0,g=0,b=0]
java.awt.Color[r=0,g=2,b=1]      java.awt.Color[r=159,g=172,b=164]      java.awt.Color[r=62,g=49,b=29]
java.awt.Color[r=68,g=54,b=33]    java.awt.Color[r=111,g=92,b=64]    java.awt.Color[r=166,g=179,b=165]
java.awt.Color[r=119,g=125,b=113]                        java.awt.Color[r=183,g=151,b=135]
java.awt.Color[r=139,g=111,b=89]                        java.awt.Color[r=167,g=137,b=115]
java.awt.Color[r=150,g=131,b=107]                        java.awt.Color[r=132,g=113,b=80]
java.awt.Color[r=156,g=124,b=102]  java.awt.Color[r=75,g=61,b=36]  java.awt.Color[r=168,g=136,b=114]
java.awt.Color[r=155,g=129,b=110]  java.awt.Color[r=125,g=110,b=79]  java.awt.Color[r=58,g=32,b=30]
java.awt.Color[r=69,g=53,b=38]      java.awt.Color[r=66,g=59,b=42]      java.awt.Color[r=97,g=107,b=100]
java.awt.Color[r=163,g=168,b=152]

95

## 5.2. Screen Shots

Following figures 9 and 10 demonstartes input and output screens. Due to limitation of space we are unable to show of intermediate results like oracle database, datatables and attribute values, etc..

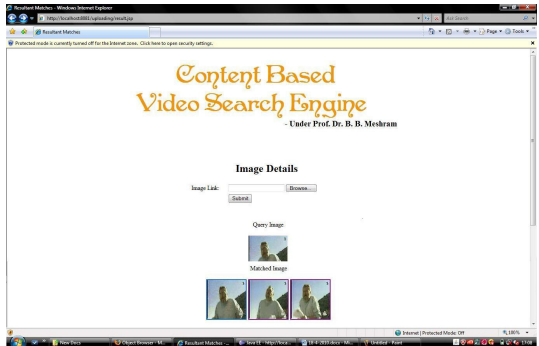 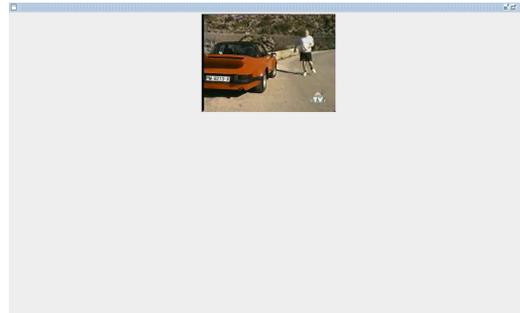

Figure 9.  Sreen  showing result of match          Figure 10.  video player maximized

## 6. CONCLUSIONS

Video Retrieval is broad area that integrates features from several features and fields including artificial intelligence, machine learning, data base management systems, etc. There have been large numbers of algorithms rooted in these fields to perform various video retrieval tasks. In our proposed system we have implemented retrieval system by integrating various features of query video frame. Experimental results show that integration of extracted features improves video indexing and retrieval. This is demonstrated by the finding that multiple features produce effective and efficient system as precision and recall values are improved. There can be large area of application of proposed system like criminal database retrieval, biomedical information, etc. We further intend to enhance system by integrating more features and other methods and can be extended further to retrieval also from social websites.

**Authors**

**B. V. Patel,** is working as Principal at Shah & Anchor Kutchhi Polytechnic, Mumbai, India. He received his B.E. in Computer Science and Engineering from Karnataka University, India, M.E. in Computer Engineering from VJTI, Mumbai, India. He has authored more than twenty papers in National and International conference and journals. He is also on editorial board of International Journals. His teaching and research interest includes Multimedia Data Mining, Network Management, Network Security, E-learning, etc. He is member of ACM, INENG, IACSIT and CSI.

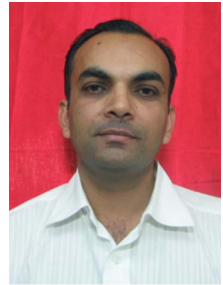

**Dr. B. B. Meshram,** is working as Professor & Head of Computer Technology Dept., VJTI, Matunga, Mumbai. He is Ph.D. in Computer Engineering and has more than 100 papers to his credit at National and International level including international journals. He is the life member of CSI and Institute of Engineers.

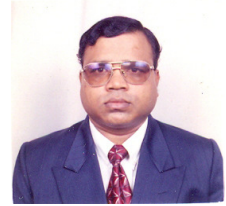